\documentclass[superscriptaddress,showpacs,amsmath,amssymb,bibnotes,footinbib,reprint,aps,prb]{revtex4-1}

\usepackage{graphicx}
\usepackage[colorlinks=true,linkcolor=blue,citecolor=blue,urlcolor=blue]{hyperref}

\begin{document}

\title{Half-quantum vortices on \texorpdfstring{$c$}{c}-axis domain walls in chiral \texorpdfstring{$p$}{p}-wave superconductors}
\author{Sarah B. Etter}
\affiliation{Institute for Theoretical Physics, ETH Zurich, 8093 Zurich, Switzerland}
\author{Wen Huang}
\affiliation{Department of Physics and Astronomy, McMaster University, Hamilton, Ontario, L8S 4M1, Canada}
\affiliation{Shenzhen Institute for Quantum Science and Engineering, Southern University of Science and Technology, Shenzhen, 518055, Guangdong, China}
\author{Manfred Sigrist}
\affiliation{Institute for Theoretical Physics, ETH Zurich, 8093 Zurich, Switzerland}
\date{\today}

\begin{abstract}
Chiral superconductors are two-fold degenerate and domains of opposite chirality can form, separated by domain walls. There are indications of such domain formation in the quasi two-dimensional putative chiral $p$-wave superconductor Sr$_2$RuO$_4$, yet no experiment has explicitly resolved individual domains in this material. In this work, $c$-axis domain walls lying parallel to the layers in chiral $p$-wave superconductors are explored from a theoretical point of view. First, using both a phenomenological Ginzburg-Landau and a quasiclassical Bogoliubov-deGennes approach, a consistent qualitative description of the domain wall structure is obtained. While these domains are decoupled in the isotropic limit, there is a finite coupling in anisotropic systems and the domain wall can be treated as an effective Josephson junction. In the second part, the formation and structure of half-quantum vortices (HQV) on such $c$-axis domain walls are discussed.
\end{abstract}

\maketitle

\section{Introduction}
The unconventional superconductivity in the quasi two-dimensional layered perovskite Sr$_2$RuO$_4$, discovered more than two decades ago\cite{maeno:1994}, continues to attract considerable interest. Following early predictions of spin-triplet superconductivity in connection with the strong Hund's coupling between the Ru $d$-orbitals\cite{rice:1995,baskaran:1996}, a large number of experiments have pointed toward an odd-parity\cite{nelson:2004,liu:2010}, spin-triplet\cite{ishida:1998,duffy:2000,anwar:2016} and time reversal symmetry-breaking\cite{luke:1998,xia:2006,kapitulnik:2009} superconducting state. This makes chiral $p$-wave pairing the most probable candidate\cite{mackenzie:2003,maeno:2012,kallin:2012}, although challenging discrepancies remain\cite{mackenzie:2017}. In particular, recent NMR-Knight shift data show discrepancies to earlier results being more consistent with spin singlet pairing \cite{pustogow:2019,ishida:2020,petsch:2020}. It is unclear so far whether spin-orbit coupling and the multi-orbital nature of the electronic band structure would make the results compatible with spin triplet symmetry \cite{ramires:2016,oda:2019}. 

The chiral $p$-wave state is two-fold degenerate with phases of positive and negative chirality $\pm$ connected through time reversal. 
If full rotation symmetry around the $z$-axis is present in the electronic structure, the corresponding Cooper pair states can be attributed a definite angular momentum $\pm \hbar$. The formation of domains of opposite chirality is possible upon the nucleation of the superconducting phase and depends on the cooling process. However, despite extensive experimental investigation no direct observation of the domains has been reported such that their structure and size is unknown to date. 

Two typical domain geometries should be considered for a layered superconductor with a quasi-2D electronic structure. The first is the $ab$-plane (inplane) domain wall which separates regions of opposite chiralities within the same layer. These domain walls support chiral quasiparticle modes indicative of the non-trivial topological nature of the chiral state\cite{read:2000,furusaki:2001}. The details of the domain wall structure, such as the stable configurations and the behavior of the two chiral components, have been studied previously within the Ginzburg-Landau theory\cite{volovik:1985,sigrist:1989,sigrist:1999}, as well as within a quasiclassical approximation\cite{matsumoto:1999,mukherjee:2015,bouhon:2014a}. The existence of these domains was proposed to explain the spontaneous internal magnetic field observed in $\mu$SR\cite{luke:1998,matsumoto:1999} and the unusual  interference patterns in extended Josephson junctions both between Sr$_2$RuO$_4$ and a conventional superconductor\cite{kidwingira:2006,bouhon:2010}, and in single crystal ring structures\cite{yasui:2020}. However, no direct observation of domain formation has been reported in real space scanning probes\cite{kirtley:2007,hicks:2010,curran:2014}. Furthermore, the size of the domains inferred from the existing measurements seems inconsistent \cite{kallin:2009}.

The other geometry is the $c$-axis domain wall which spans the $ab$-plane. This type of domain wall is less studied. However, it is energetically more favorable in comparison to the inplane domain walls owing to the weak interlayer coupling. Experimentally, it has been proposed that $c$-axis domains may explain the observed absence of the spontaneous flux at the surface\cite{hicks:2010}.

In the present work we discuss the structure and magnetic properties of $c$-axis domain walls. In the isotropic limit (full rotation symmetry around the $c$-axis), such domain walls decouple the chiral domains completely from each other, because the angular momentum of the Cooper pairs is a good quantum number and restrictive selection rules apply in Cooper pair tunneling. Thus, a supercurrent along the $c$-axis through the domain wall would not be possible. 
By introducing anisotropy into the electronic band structure, however, a finite coupling appears, and supercurrents can flow between the chiral domains. Here we present a comprehensive discussion of the $c$-axis domain wall from both a phenomenological Ginzburg-Landau and a quasiclassical Bogoliubov-deGennes view point. Both descriptions confirm the finite coupling away from the isotropic limit, and find that the phase shift across the junction depends on the sign of the anisotropy and exhibits a non-trivial periodicity of $\pi$. 
Like Josephson junctions such domain walls can host vortices which due to the $ \pi $-periodicity of the phase are 
half quantum vortices (HQV). We note that the HQVs considered here belong to the class of fractional vortices within the framework of multicomponent order parameters\cite{sigrist:1991} and are fundamentally different from the HQVs based on the spin rotation of the Cooper pairs in spin-triplet superconductors which have attracted a lot of interest for their non-trivial topological properties\cite{kopnin:1991,ivanov:2001,nayak:2008}. We show that the HQVs on the $c$-axis domain wall can indeed be stable, and that both the maximal current across the domain wall and the characteristic length scales of the HQV depend on the magnitude of the anisotropy. By tuning the system towards the isotropic limit, where the two domains are decoupled, the HQV dissolves along the domain wall. For stronger anisotropies, when the HQV becomes smaller than the relevant screening length, non-local effects have to be taken into account.

In the following, first, the phenomenological Ginzburg-Landau approach is presented in Sec.~\ref{sec:gl}, providing both an approximative analytical and a self-consistent numerical solution. This is complemented by a quasiclassical Bogoliubov-deGennes approach in Sec.~\ref{subsec:BdG}. Next, we explore the HQV on the domain wall. Its general structure is described in Sec.~\ref{sec:struc}, while the full junction phenomenology is analyzed in Sec.~\ref{sec:sg} within a sine-Gordon framework for both the isotropic and the non-local limit.

\section{Domain wall structure}

We start with the analysis of the basic structure of $c$-axis domain walls in spin-triplet chiral $p$-wave superconductors for a systems with an anisotropic electronic structure. 
For a system with full rotation symmetry around the $c$-axis this state possesses a definite angular momentum $ L_z = \pm 1 $ with a gap function $ {\boldsymbol d} = \Delta_0 \hat{\boldsymbol z} (k_x \pm i k_y) $. Much of the phenomenology of this pairing state
can be transferred to other chiral superconducting phases with the same angular momentum property, such as the spin-singlet chiral $d$-wave state with the gap function 
$ \psi ({\boldsymbol k}) =   \Delta_0 k_z (k_x \pm i k_y)$.  While we present here our discussion for the chiral $p$-wave state, all qualitative results also apply to related chiral states. 

\subsection{Phenomenological Ginzburg-Landau approach}
\label{sec:gl}

The Ginzburg-Landau (GL) theory allows for a very efficient symmetry based approach to inhomogeneous structures of a superconducting  order parameter. It will provide us with the essential ingredients for the study of the HQV in the second part of this paper, Sec.~\ref{sec:hqv}.

The GL free energy of chiral $p$-wave superconductors is constructed from a two-component order parameter belonging to the two-dimensional irreducible representation $E_u$ of the full tetragonal point group $D_{4h}$, as used to describe the odd-parity spin-triplet pairing state given by $\boldsymbol d(\boldsymbol r,\boldsymbol k)= \hat{\boldsymbol z}\left(\eta_x(\boldsymbol r) f_x(\boldsymbol k)+\eta_y(\boldsymbol r) f_y(\boldsymbol k)\right)$ in the d-vector notation, where by $ {\boldsymbol d} \parallel \hat{\boldsymbol z} $ corresponds to spin configuration of inplane equal-spin pairing \cite{sigrist:1991}. Here $\{f_x(\boldsymbol k),f_y(\boldsymbol k)\}$ are basis functions of $ E_u $, odd in $ \boldsymbol k $, and $\boldsymbol \eta=(\eta_x,\eta_y)$ denotes the two-component order parameter which in the bulk takes the form $\boldsymbol\eta=\eta_\mathrm{b}(1,\pm i)$ with the two chiralities $\pm$. 

The GL free energy functional is a scalar under all symmetry operations and, thus, given by\cite{sigrist:1991}
\begin{align}
&\mathcal{F}[\eta_x,\eta_y,\boldsymbol{A}]=\int_{V_p}\mathrm{d}^3r \Big[a\left(T-T_c\right)|\boldsymbol{\eta}|^2\\
&\quad+b_1|\boldsymbol{\eta}|^4+\frac{b_2}{2}\left(\eta_x^{*2}\eta_y^2+\eta_x^2\eta_y^{*2}\right)+b_3|\eta_x|^2|\eta_y|^2\nonumber\\
&\quad+K_1\left(|D_x\eta_x|^2+|D_y\eta_y|^2\right)+K_2\left(|D_x\eta_y|^2+|D_y\eta_x|^2\right)\nonumber\\
&\quad+\Big\{K_3(D_x\eta_x)^*(D_y\eta_y)+K_4(D_x\eta_y)^*(D_y\eta_x)+\mathrm{c.c}\Big\}\nonumber\\
&\quad+K_5\left(|D_z\eta_x|^2+|D_z\eta_y|^2\right)+{\left(\nabla\times\boldsymbol{A}\right)^2}/{(8\pi)}\Big],\nonumber
\end{align}
where $V_p$ is the superconducting region, $T_c$ the critical temperature, $\{a, b_i, K_i\}$ the GL coefficients, $\boldsymbol A$  the vector potential, and $\boldsymbol D=\nabla-i\gamma\boldsymbol A$  the gauge-invariant derivative, where $\gamma=2\pi/\Phi_0$ with $\Phi_0$ the flux quantum. We do not resolve the individual RuO$_2$-layers along the $c$-axis as in a Lawrence-Doniach type of model of weak interlayer coupling\cite{kogan:1981,blatter:1992,chapman:1995}, because in Sr$_2$RuO$_4$ the coherence length $\xi_c$ along the $c$-direction is considerably longer than the interlayer distance $s$ \cite{mackenzie:2003}.

\subsubsection{Anisotropy and GL coefficients}

The GL expansion coefficients $\{a,b_i,K_i,K_5\}$ are material dependent parameters and can either be extracted from the corresponding microscopic Hamiltonian or experimental data. Their range is subject to the condition of stability of the free energy\cite{sigrist:1991}. Relations between the sets $a$, $\{b_i\}$, $\{K_i\}$ and $K_5$ can easily be determined from the linearized GL equations through the experimentally measured coherence lengths $\xi_{ab}$ and $\xi_{c}$, the London penetration depths $\lambda_{ab}$ and $\lambda_c$, and the GL parameter $\kappa$. This information is useful to write the free energy in dimensionless units such that the only experimental quantities entering are the superconducting anisotropy $\gamma_s=\xi_{ab}/\xi_c$ and the GL parameter $\kappa\equiv\kappa_{ab} = \lambda_{ab} / \xi_{ab} $, which for Sr$_2$RuO$_4$ are 20 and 2.6, respectively\cite{maeno:2012}.

The ratios within the sets $\{b_i\}$ and $\{K_i\}$, on the other hand, originate from further details of the electronic structure\cite{etter:2017}. In a weak-coupling approach for the chiral $p$-wave state on a single band (e.g. $\gamma$-band of Sr$_2$RuO$_4$) the coefficients $b_i$ are related through the band (Fermi surface) and gap structure as
\begin{subequations}
\begin{align}
b_1 &\sim  \langle f_x^4  \rangle_{FS} \\
b_2 &\sim 2\langle f_x^2 f_y^2 \rangle_{FS}\\
b_3 & =2(b_2-b_1),\label{eq:b3}
\end{align}
\end{subequations}
where $\langle \cdot \rangle_{FS}$ denotes the average over the Fermi surface.
Introducing the anisotropy parameter of the electronic structure as\cite{agterberg:1998}
\begin{equation}
\nu=\frac{\langle f_x^4 \rangle_{FS}-3\langle f_x^2 f_y^2 \rangle_{FS}}{\langle f_x^4 \rangle_{FS}+\langle f_x^2 f_y^2\rangle_{FS} }
\end{equation}
with  $\nu\in ]-1,1[$, these further reduce to
\begin{subequations}
\begin{align}
b_1 &=\frac{3+\nu}{8}b\\
b_2 &=\frac{1-\nu}{4}b \\
b_3 &= - \frac{3 \nu +1}{4} b .
\end{align}
\end{subequations}
For the sake of definiteness we use $f_x(\boldsymbol k)= v_x(\boldsymbol k)$ and $f_y(\boldsymbol k)=v_y(\boldsymbol k)$, with $v_i$ the components of the Fermi velocity, such that the anisotropy of the gap function is identified with the anisotropy of the Fermi surface\cite{bouhon:2014a}. The constant $b=2b_1+b_2$ is chosen such that the amplitude of the bulk order parameter is given by $|\eta_\mathrm{b}(T)|^2=-a(T-T_c)/b$.

The parameter $ \nu $ is a measure for the anisotropy of superconducting phase as imposed by the electron band and gap structure\cite{bouhon:2014a}. In the isotropic limit, i.e. for a completely rotationally symmetric system with a cylindrical Fermi surface, the basis functions are $f_x(\boldsymbol k) \propto k_x$ and $f_y(\boldsymbol k) \propto k_y$, resulting in $\nu=0$. Note that $ \nu = \pm 1 $ corresponds to placing the free energy at the boundary of the stable region of the chiral $p$-wave state \cite{sigrist:2005}. 

The analogous discussion for the coefficients of the gradient terms, $\{K_i\}$, leads to $K_1=(3+\nu)/4 K$ and $K_2=K_3=K_4=(1-\nu)/4 K$ with the constant $K=K_1+K_2$. These inplane gradient coefficients expressed in terms of the anisotropy $\nu$ will only enter our discussion in the later part on the HQV, Sec.~\ref{sec:hqv}. For the structure of the $c$-axis domain wall as discussed below, the order parameter is translationally invariant for inplane coordinates. 

\subsubsection{Phase shift across the domain wall}

Domain walls involve a spatial change of the order parameter which can be decomposed into a variation of the amplitude of the order parameter components in the two domains and a shift of the overall phase across the domain wall. 
First we discuss which phase shift minimizes the free energy, depending on the anisotropy of the electronic structure, and then address the shape of the order parameter across the domain wall in the following section.

To analyze the domain wall structure it is convenient to use an order parameter representation $\boldsymbol\eta=(\eta_+,\eta_-)=\big((\eta_x-i\eta_y)/2, (\eta_x + i\eta_y)/2\big)$ directly addressing the two degenerate chiral states. It is further useful to parametrize the complex order parameter components in terms of amplitude and phase $\eta_\pm=|\eta_\pm|e^{i\phi_\pm}$. With the choice of gauge $\gamma A_z=\partial_z\phi_+$ only the relative phase $\varphi=\phi_-(z>z_{dw})-\phi_+(z<z_{dw})$ enters the free energy, with $ z_{dw} $ the position of the domain wall leading to an order parameter of the form $ {\boldsymbol \eta} = ( | \eta_+| , |\eta_-| e^{i\varphi} $),
\begin{align}
\begin{split}
&\mathcal{F}[\eta_+,\eta_-]=\int\mathrm{d}z\Big[2a(T-T_c)(|\eta_+|^2+|\eta_-|^2)\\
&\quad+(2b_1+b_2)\left(|\eta_+|^4+4|\eta_+|^2|\eta_-|^2+|\eta_-|^4\right)\\
&\quad+(4b_1-6b_2)\left(|\eta_+|^2|\eta_-|^2\cos\left(2\varphi\right)\right)\\
&\quad+2K_5\big((\partial_z|\eta_+|)^2+(\partial_z|\eta_-|)^2+|\eta_-|^2 \left(\partial_z\varphi\right)^2\big)\Big].
\end{split}
\end{align}
Here we neglect variations of the order parameter phase for $ |z-z_{dw}| \gg \xi' $, where one order parameter component vanishes, with $ \xi' $ the width of the domain wall (we use $ z_{dw} = 0$). As the basic structure of the domain wall only depends on the out-of-plane $z$-direction, the inplane spatial dependence can be neglected completely.

It is immediately obvious that the free energy is minimized for $\varphi=\mathrm{const}$, such that $\partial_z\varphi=0$. Therefore we 
may model  the $c$-axis domain wall using the boundary conditions
\begin{subequations}\label{eq:bcpm}
\begin{align}
(\eta_+,\eta_-)&=(|\eta_\mathrm{b}|,0)\quad(z\rightarrow-\infty)\\
(\eta_+,\eta_-)&=(0,|\eta_\mathrm{b}| e^{i\varphi})\quad(z\rightarrow\infty),
\end{align}
\end{subequations}
with the bulk value $\eta_\mathrm{b}$ given above. Straightforward symmetry considerations lead to the general structure $\boldsymbol\eta=(g(z),g(-z)e^{i\varphi})$, with $g(z)$ being real, positive and asymmetric with respect to the domain wall ($ g(z) \to 0 $ for $ z \to + \infty $ and $ g(z) \to |\eta_b| $ for $ z \to - \infty $). The only remaining term depending on the phase shift $ \varphi $ is then
\begin{align}
\begin{split}
F_\varphi &= 2(2b_1-3b_2) |\eta_+|^2 |\eta_-|^2 \cos(2\varphi)\\
&=2\nu b (g(z) g(-z))^2 \cos(2\varphi).
\label{eq:f_phi}
\end{split}
\end{align}
Note that this term is concentrated on the domain wall region and vanishes for $ | z | \gg \xi' $. 
The \emph{sign} of the anisotropy of the electronic structure $\nu$ thus single-handedly determines the most energetically favorable choice of $\varphi$ across the domain wall, 
\begin{equation}\label{eq:varphi}
\varphi=\begin{cases}     
0\mod \pi & \nu < 0 \\
\forall & \nu=0 \\
\pi/2\mod\pi & \nu>0.
\end{cases}
\end{equation}
The resulting free energy and, thus, the shape of the order parameter, on the other hand, only depend on the \emph{magnitude} of the absolute value $|\nu|$. 

We would like to comment here on our approximation to reach the solution for $ \varphi $ in Eq.(\ref{eq:varphi}). First, we assumed that the coherence length along the $c$-axis covers many layers. If this is not the case and $ K_5 $ would be very small, the approximation, $ \partial_z \varphi = 0 $ is not justified. Then passing with $ \nu $ through zero may not result in a discontinuous change from 0 to $ \pi/2 $. Rather $ \varphi $ could move continuously between these two minima for a certain range of $ \nu $ around 0, because the $ \varphi $-dependence of the free energy would be rather weak. Indeed our numerical treatment points towards such a behavior. However, for the sake of simplicity we do not analyze this rather special limit here, as it lies outside our scope. 

In the isotropic limit ($\nu=0$), the free energy is fully degenerate for all $ \varphi $, signaling the complete phase decoupling of the two domains, analogous to the Josephson coupling between two condensates. 
This limit describes a system with complete rotation symmetry around the $c$-axis, where the Cooper pair orbital angular momentum is a good quantum number and is conserved during the tunneling event. Pair tunneling is, thus, prohibited by this selection rule, consistent with our phenomenological result. 
Once the rotational symmetry is broken the selection rule is no longer valid and Cooper pairs can be transfered, in the present case 
on the level of a second order coupling, i.e. even numbers of Cooper pairs pass together. In this way the two domains are phase coupled and a supercurrent can flow across the domain wall. 
Given the non-trivial $\pi$-periodicity of the discrete allowed values for $ \varphi $, realized through the $ \cos (2 \varphi ) $ term in the 
free energy, we will explore the possibility of HQVs on the domain wall in Sec.~\ref{sec:hqv}.

\subsubsection{Shape of the order parameter across the domain wall}

While the phase shift across the domain wall only depends on the sign of the anisotropy, the change in the amplitude of the order parameter components only depends on its magnitude $|\nu|$. We will now tackle the domain wall problem first with an approximate variational solution, which we then compare with the exact numerical solution for the shape of the order parameter. 

Following the approximations introduced in Ref.~[\onlinecite{sigrist:1999}], the total amplitude of the order parameter is kept constant everywhere $|\boldsymbol\eta|^2=|\eta_\mathrm{b}|^2$ with the ansatz $(\eta_+,\eta_-)=\eta_\mathrm{b}\left(\sin(\chi(z)),\cos(\chi(z))e^{i\varphi}\right)$. The boundary condition from Eq.~(\ref{eq:bcpm}) translates into $\chi(z\rightarrow\infty)\rightarrow0$ and $\chi(z\rightarrow-\infty)\rightarrow\pi/2$. In this way the free energy simplifies to
\begin{align}
\begin{split}
&\mathcal{F}[\eta_+,\eta_-,\varphi]=\int\mathrm{d}z\Bigg[2a(T-T_c)\eta_\mathrm{b}^2+b\eta_\mathrm{b}^4\\
&\quad+\frac{b\eta_\mathrm{b}^4}{2}\left(1-|\nu|\right)\sin^2(2\chi)+2K_5\eta_\mathrm{b}^2(\partial_z\chi)^2\Bigg],
\end{split}
\end{align}
where the phase shift $ \varphi $ given in Eq.~(\ref{eq:varphi}) has already been implemented. Therefore this free energy only depends on the absolute value of the anisotropy $|\nu|$ and not on its sign. Variation with respect to $\chi$ leads to the sine-Gordon type differential equation
\begin{align}
\partial_z^2\chi(z)=\frac{b\eta_\mathrm{b}^2\left(1-|\nu|\right)}{4K_5}\sin(4\chi(z)).
\end{align}
Using $\xi_c=\sqrt{K_5/(b\eta_\mathrm{b}^2)}$, the standard solution is
\begin{align}\label{eq:chi}
\begin{split}
\chi(z)&=\arctan\left(\exp\left(-\sqrt{1-|\nu|}\frac{z}{\xi_c}\right)\right),
\end{split}
\end{align}
such that the width of the domain wall scales as $ \xi' = 2\xi_c/\sqrt{1-|\nu|}$.
The free energy density per unit inplane area is then found by inserting this solution,
\begin{align}\label{eq:dw_en}
\mathcal{F}-\mathcal{F}_0=\frac{3}{\sqrt{2}}b\eta_\mathrm{b}^4\xi_c\sqrt{1-|\nu|},
\end{align}
where $\mathcal{F}_0$ is the bulk free energy without the domain wall.

The cost of the domain wall is largest in the isotropic limit $|\nu|=0$, where the two domains are fully phase decoupled. 
On the other hand, at the stability boundary of the chiral $p$-phase, $ \nu = \pm 1 $, the domain wall energy vanishes and the width $ \xi' $ diverges.
Compared to the inplane domain wall whose width is connected with $\xi_{ab}$, the $c$-axis domain wall scales with $\xi_c $ ($\ll\xi_{ab}$) and is energetically much cheaper. 

We minimize the free energy numerically setting $T=0$ in the free energy and without any restrictions on the amplitude of the order parameter. Fixed values are used for $ \varphi $, however, according to Eq.~(\ref{eq:varphi}), $\varphi=\{0,\pi/2\}$. The numerical scheme follows a one-step relaxed Newton-Jacobi method for boundary value problems\cite{gardan:1985,ortega:2000,piette:2004}, details see Ref.~[\onlinecite{etter:2017}].

\begin{figure}
\includegraphics[width=0.9\columnwidth]{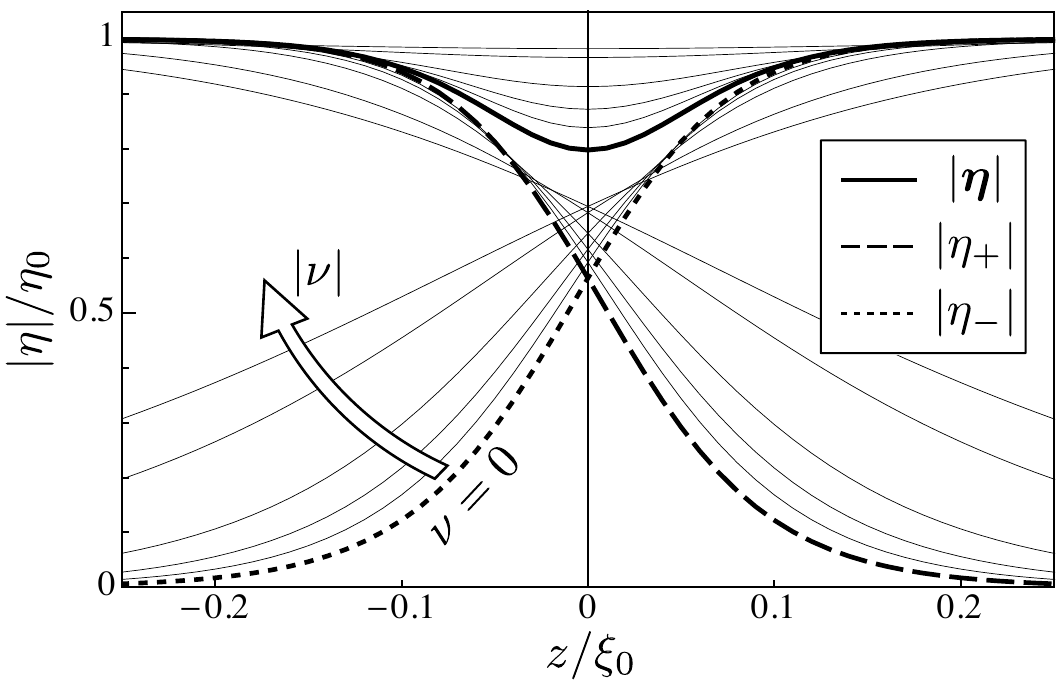}
\caption{The shape of the order parameter amplitudes across the $c$-axis domain wall for different anisotropies $|\nu|\in\{0,0.3,0.5,0.7,0.9,0.95\}$ resulting from a numerical minimization. The shape only depends on the magnitude of the anisotropy $|\nu|$. For $|\nu|\rightarrow1$, the domain wall dissolves and its width diverges. The total absolute value $|\boldsymbol\eta|$ is suppressed at the domain wall, the most for $\nu=0$.}
\label{fig:op}
\end{figure}

The numerical results for the two components $|\eta_+|$ (dashed) and $|\eta_-|$ (dotted) are shown in Fig.~\ref{fig:op} together with the total absolute value $|\boldsymbol\eta|$ (solid curve), all at $\nu=0$. In addition, the results for higher values of $\nu$ are indicated by the thin lines which display the growing domain wall width for increasing $ | \nu | $. Unlike in our variational ansatz the value of $ |\boldsymbol \eta| $ shows a dip at the domain wall which is weakened as $ | \nu | $ grows and is entirely constant in the limit $ |\nu| \to 1 $. 
Finally, we note that the symmetry $(\eta_+,\eta_-)\propto\left(g(z),g(-z)\right)$ is borne out in the numerical solution.

\begin{figure}
\includegraphics[width=0.9\columnwidth]{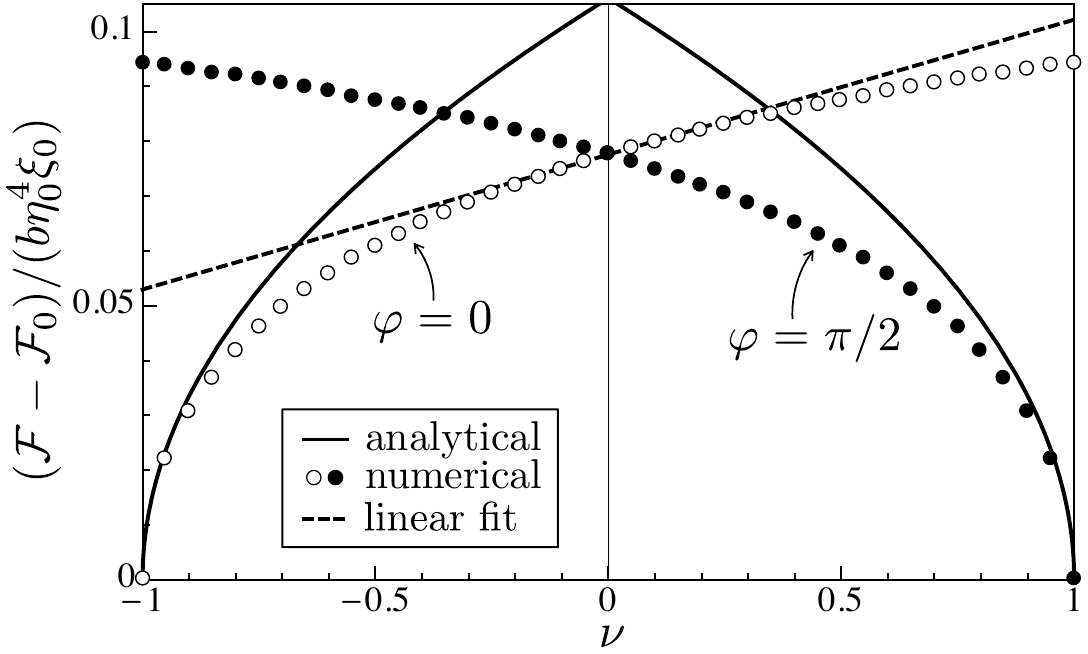}
\caption{Free energy density per unit area of the $c$-axis domain wall comparing the approximative analytical solution (solid black line) with the numerical result for $\varphi=0$ (empty dots) and $\varphi=\pi/2$ (filled dots), minimizing the free energy for $\nu<0$ and $\nu>0$, respectively. A linear fit to the numerical data for small $|\nu|\approx0$ (dashed) is also indicated.}
\label{fig:fe}
\end{figure}

The numerical result for the free energy density is shown in Fig.~\ref{fig:fe} for $\varphi=\{0,\pi/2\}$ (empty and filled dots), together with the approximative analytical solution from Eq.~(\ref{eq:dw_en}) (solid black line), as a function of the anisotropy $\nu$. In addition, a linear fit to the numerical data for $\nu\approx0$ is indicated (dashed line), which will be used in Sec.~\ref{sec:sg}.
The crossing of the two energy branches at $ \nu =0 $ indicates a first order change for $ \varphi $ between the two sectors of different values as given in Eq.~(\ref{eq:varphi}).
The qualitative agreement between the numerical solution and the variational approximation is very good. Close to $ | \nu | \to 1 $ even a quantitative agreement is found, as in this limit the dip in $ | \eta | $ disappears in the numerical solution, in accordance to the simplifying assumption for the variational ansatz above. Finally we note the symmetry $\mathcal{F}[\nu,\varphi]=\mathcal{F}[-\nu,\varphi+\pi/2]$ always holds. 

\subsection{Quasiclassical Bogoliubov-deGennes approach}
\label{subsec:BdG}

In this section, we address the $c$-axis domain wall from a more microscopic viewpoint through a self-consistent Bogoliubov-deGennes (BdG) treatment. 
For simplicity, we restrict to a one-band spinless Fermion tight-binding model on a square lattice which is sufficient for a spin-triplet superconductor. 
The conclusions are, however, applicable to spinful and multi-band systems. 
In contrast to the GL analysis above, here the vector potential is neglected. Nevertheless, the solutions obtained show the same main behavior and give an insight on the role of the quasiparticle states at the domain wall. 

To simulate a $c$-axis domain wall in numerical BdG, we also use inplane translational invariance in a system of $N_z$ layers. The corresponding mean-field BdG formulation is given by
\begin{align}
H = \sum_{\boldsymbol k}\sum_{l=-N_z/2}^{+N_z/2}  & \Big( \xi_{\boldsymbol k} c^\dagger_{l,\boldsymbol k} c_{l,\boldsymbol k} -t_z \sum_{\boldsymbol \delta = \pm 1 }c^\dagger_{l+\boldsymbol \delta,\boldsymbol k} c_{l,\boldsymbol k}   \nonumber \\
&  + \eta_{l,\boldsymbol k} c^\dagger_{l,\boldsymbol k} c^\dagger_{l,-\boldsymbol k} + \eta^\ast_{l,\boldsymbol k} c_{l,-\boldsymbol k} c_{l,\boldsymbol k}  \Big), &&
\end{align}
where $c^\dagger_{l,\boldsymbol k}$ ($c_{l,\boldsymbol k}$) creates (annihilates) an electron in the $l$-th layer with inplane momentum $\boldsymbol k$ and $\xi_{\boldsymbol k} = -2 t(\cos k_x + \cos k_y) - 4 t^\prime \cos k_x \cos k_y -\mu$ denotes the inplane dispersion relation. The last two terms represent the superconducting pairing with chiral $p$-wave symmetry, $\eta_{l,\boldsymbol k} = \eta_{x}(l)f_{x}(\boldsymbol k) \pm i \eta_{y}(l) f_{y}(\boldsymbol k)$, where we will use two types of pairing states assuming nearest neighbor [$ (f_x,f_y) =  (\sin k_x, \sin k_y)$]  or next-nearest neighbor [$ (f_x,f_y) = (\sin k_x \cos k_y, \sin k_y \cos k_x)$]  pairing interactions. Note the quasi-two-dimensional band structure of Sr$_2$RuO$_4$ implies that the dispersion along the $c$-axis is very small, $t_z \ll t,t^\prime$. Without loss of generality, $t=1$, $t^\prime=0.375t$ and $t_z=0.03t$ are used throughout this section.

The $c$-axis domain wall is formed by allowing the two chiral components $\eta_x$ and $\eta_y$ to vary from layer to layer. Far from the domain wall the pairing states on the two sides correspond to the bulk states of opposite chirality,
\begin{equation}
  \begin{cases} 
      \hfill e^{-i\frac{\varphi}{2}} \eta_{\boldsymbol k,+}= e^{-i\frac{\varphi}{2}} \Delta_0 [f_x(\boldsymbol k) + i f_y (\boldsymbol k)] ~,    \hfill &  l = -N_z/2 \\
      \hfill e^{+i\frac{\varphi}{2}} \eta_{\boldsymbol k,-} = e^{+i\frac{\varphi}{2}}  \Delta_0 [f_x(\boldsymbol k) - i f_y (\boldsymbol k)]~, \hfill & l = +N_z/2  \\
  \end{cases}
\label{eq:Domain0}
\end{equation}
with an additional phase shift $\varphi$ imposed between the two domains and $ \Delta_0 $ the self-consistently determined bulk value of gap function amplitude. To perform the self-consistent calculation of the bulk and the domain wall structure for the two cases of pairing states we introduce either a purely nearest-neighbor or a purely next-nearest-neighbor pairing interaction. We start the iteration process for self-consistency with an initial configuration for a domain wall at the center of the system ($l=0$), following Ref.~[\onlinecite{sigrist:1999}],
\begin{eqnarray}
\eta(l) &=& \eta_+ (l)+ \eta_-(l) \nonumber \\
&=& \eta_{\boldsymbol k,+} e^{-i\frac{\varphi}{2}} \cos \chi(l) + \eta_{\boldsymbol k,-}e^{+i\frac{\varphi}{2}}\sin \chi(l)  \,,
\label{eq:Domain}
\end{eqnarray}
where $\chi(l)= \frac{\pi}{4}(1+\tanh\frac{l}{\lambda})$, with $\lambda$ being a constant that defines the initial value for the extension of the domain wall along $z$. The structure of this initial configuration may or may not correspond to an actual energetically favorable solution. In the case of the latter, the system in general evolves into a stable state after sufficient iterative steps in our self-consistence procedure. For the calculations in Figs.~\ref{fig:DomainEnergy}, $\lambda=2$ was used in the initial configuration.

\subsubsection{Stable domain wall configurations}
\begin{figure}
\includegraphics[width=0.9\columnwidth]{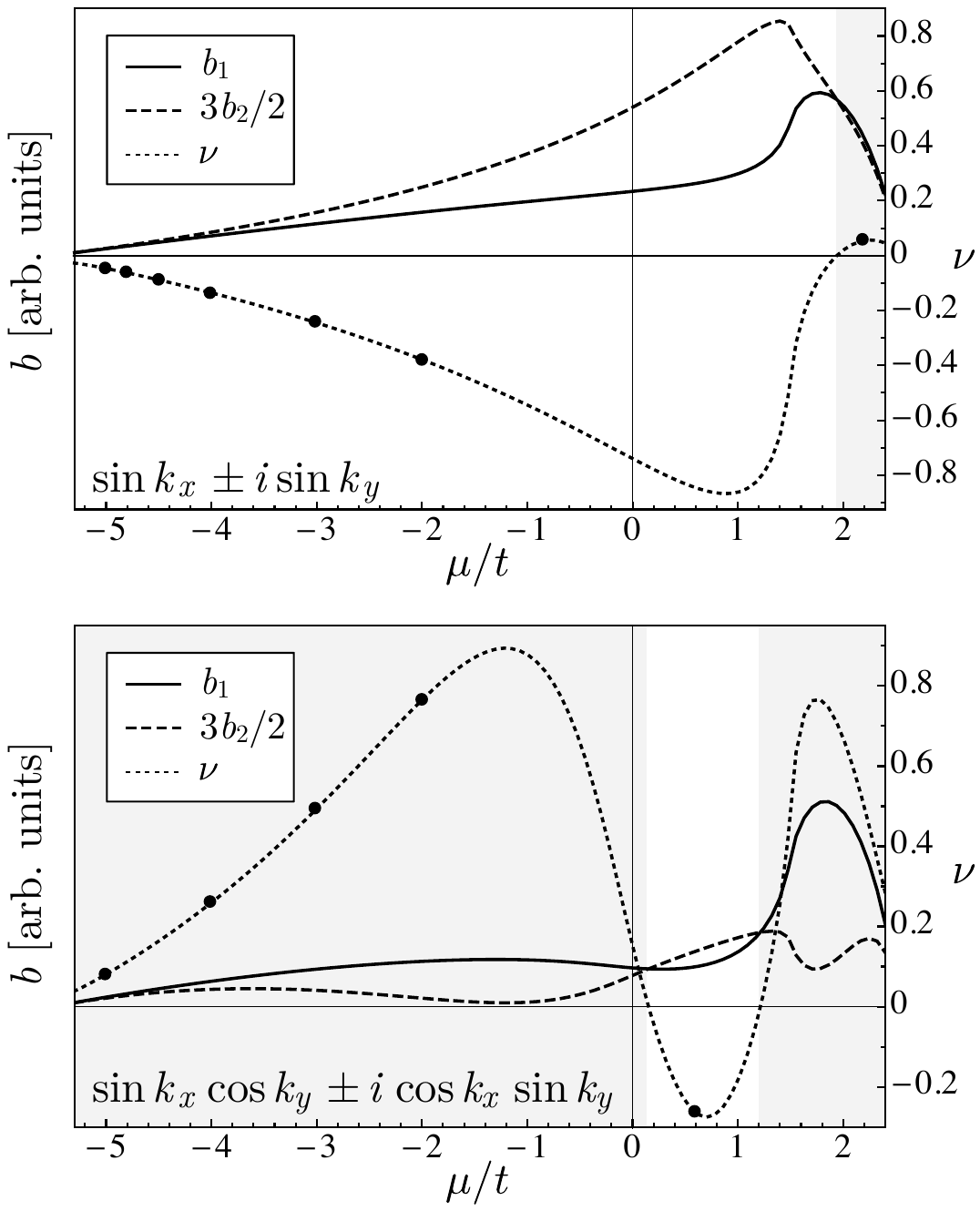}
\caption{Behavior of the GL parameters $b_1$ (solid), $3b_2/2$ (dashed) and of the anisotropy parameter $\nu$ (dotted) as a function of chemical potential for both NN-pairing (top) and NNN-pairing (bottom) on a square lattice. Regions of positive anisotropy are shaded gray. Black dots indicate the examples shown in Fig.~\ref{fig:DomainEnergy}. Note that results near the continuum limit at the bottom and top of the band are excluded due to poor convergence of the integration in our implementation. Nonetheless, in this limit $b_1 \rightarrow 3b_2/2$ and $\nu \rightarrow 0$, as expected. For the calculations we used the parameters for hopping $ t' = 0.375 t $, $ t_z = 0.03 t $ and the gap amplitude $ \Delta_0 = 0.1 t $ where $ t $ defines the unit of energy.}
\label{fig:nu}
\end{figure}

In lattice models, the anisotropy parameter $\nu$ is generically non-vanishing and shall depend on the details of the gap and band structure anisotropy.  In Fig.~\ref{fig:nu} we plot the relevant parameters as a function of chemical potential for two different chiral $p$-wave gap functions on a square lattice. Consistent with the GL treatment, our variational BdG yields two sectors of stable domain walls, depending on the sign of $\nu$. 

\begin{figure}
\includegraphics[width=0.9\columnwidth]{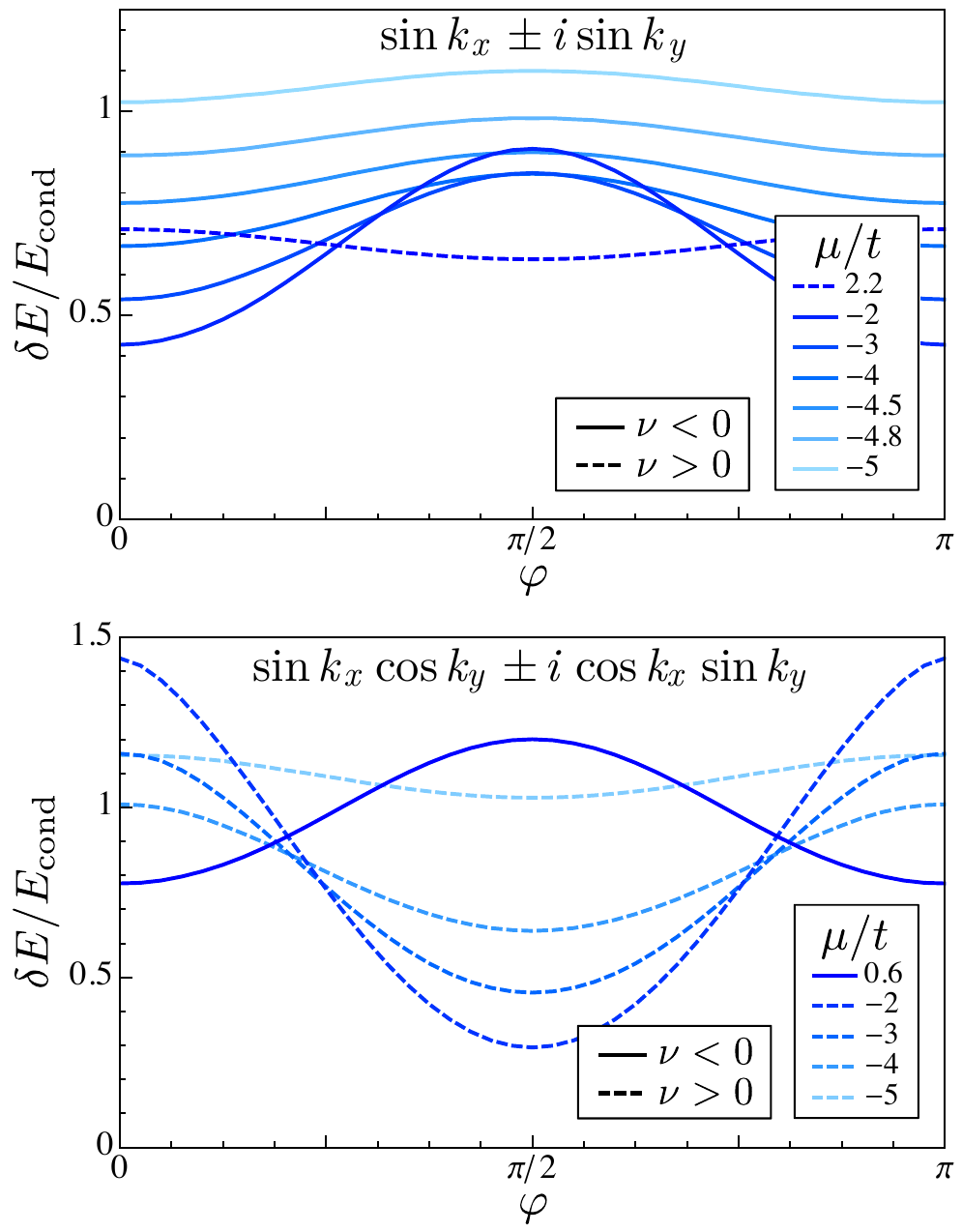}
\caption{Non-self-consistent domain wall energy as a function of $\varphi$ with respect to the condensation energy density per $ab$-layer for a range of chemical potentials and for both NN- and NNN-pairings. The calculations were performed on the same model as in Fig.~\ref{fig:nu}, where the values of $\nu$ are indicated by black dots.}
\label{fig:DomainEnergy}
\end{figure}

Next we would like to understand the role of the phase shift $ \varphi $. For this purpose we use a non-self-consistent approach of the BdG scheme where the gap function is taken
with the fixed  $\chi(l)$ as given in Eq.~(\ref{eq:Domain}) while keeping $ \varphi $ as a free parameter. In this way we deduce the energy of the domain wall as a function of $ \varphi $, relative to the bulk condensation energy which is approximated by 
\begin{equation}
E_\text{cond}= \frac{1}{4\pi^2}\left\langle\frac{|\eta_{0\boldsymbol k}|^2}{\sqrt{v_{x,\boldsymbol k}^2+v_{y,\boldsymbol k}^2}} \right\rangle_{FS}
\end{equation}
per layer. 
Fig.~\ref{fig:DomainEnergy} displays the energies for the types of gap functions introduced above (NN-pairing in the upper and NNN-pairing in the lower panel) for varying chemical potential. The anisotropy $ \nu $ corresponding to the different curves connect with the black dots given in Fig.~\ref{fig:nu}. The minima are found at $ \varphi =0,\pi $ for negative $ \nu $ and at $ \varphi = \pi/2 $ ($3 \pi /2 $) for positive $ \nu $. The smaller $ | \nu | $ the weaker the $ \varphi $-dependence, as expected from our previous discussion. 

\subsubsection{Domain wall Andreev quasiparticle states}
The emergence of subgap chiral quasiparticles at the $ab$-plane domain walls is a well-known feature whose origin lies in the topological nature of the superconducting phase \cite{matsumoto:1999,furusaki:2001}.  The $c$-axis domain walls also host localized subgap quasiparticle states which, in contrast, do not directly reflect topological properties. 
Nevertheless, the chiral pairing does have its impact on the spectrum of these subgap Andreev states and their contribution to the coupling of the two domains. 

\begin{figure}
\includegraphics[width=0.5\columnwidth]{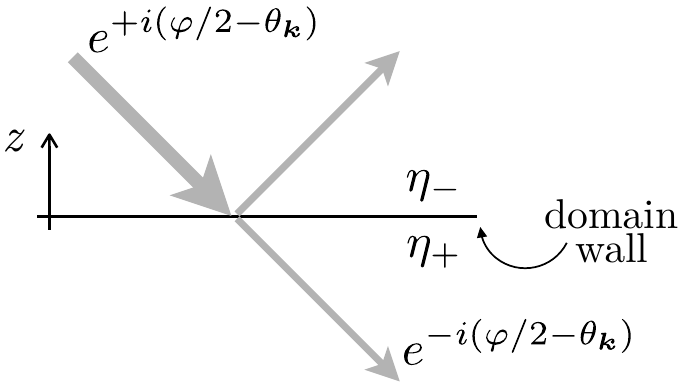}
\caption{Trajectories of quasiparticle waves at the $c$-axis domain wall. The quantity $\varphi$ indicates the phase of the superconducting order parameter in the two domains, while $\theta_{\boldsymbol k}$, where ${\boldsymbol k}$ is the inplane wavevector, denotes the direction of the wavevector. The inplane component of the momentum is conserved in the scattering process.}
\label{fig:Semiclassical}
\end{figure}

For our analysis we consider the situation from the viewpoint of a planar $c$-axis Josephson junction where the two connected superconductors are chiral, for simplicity described by a gap function on the Fermi surface of the form $ \eta_{\boldsymbol k, \pm} = \eta_0 e^{\pm i \theta_{\boldsymbol k}} $. We examine the electronic states based on a quasiclassical approach, focussing on particle trajectories between the two superconductors (see Fig.~\ref{fig:Semiclassical}). Such a trajectory can be labelled by the momentum $ \boldsymbol k = (k \cos \theta_{\boldsymbol k}, k \sin \theta_{\boldsymbol k}, k_z ) $, near the Fermi surface. Following Fig.~\ref{fig:Semiclassical} the trajectory connects gap functions of the phase  $\varphi/2-\theta_{\boldsymbol k}$ in $ \eta_- $-domain with those of $ - \varphi/2+ \theta_{\boldsymbol k}$ in the $ \eta_+ $-domain. Consequently, for such a trajectory the corresponding phase difference between the two domains is $ \varphi - 2 \theta_{\boldsymbol k} $. 
The set of quasiparticle states derived from each trajectory (including the reflected part) contributes to the coupling and the associated energy depends on the phase difference, $ E (\varphi - 2 \theta_{\boldsymbol k},\theta_{\boldsymbol k}) $ and, in general, on the direction $ \theta_{\boldsymbol k} $. The total energy of the junction is the integral over all trajectories, here reduced to the angles $ \theta_{\boldsymbol k} $, as we restrict to momenta at the Fermi surface,
\begin{equation}
E_{\mathrm{tot}} (\varphi) = \int_0^{2\pi} d \theta \; E(\varphi- 2 \theta,\theta) .
\label{e-tot}
\end{equation}
Together with the standard periodicity $ E_{\mathrm{tot}} (\varphi) = E_{\mathrm{tot}} (\varphi + 2 \pi n) $ this corresponds well to $ \cos (2 \varphi ) $ in lowest order coupling. 
This simplified viewpoint allows us now to give a qualitative discussion of the role of anisotropy. For full rotation symmetry, $ E(\tilde{\varphi} , \theta) = E(\tilde{\varphi}, \theta + \alpha) $ for an arbitrary angle $ \alpha $. Thus, $ E(\varphi-2 (\theta +\alpha) , \theta + \alpha) = E(\varphi'- 2\theta,\theta) $ with $ \varphi'=\varphi- 2 \alpha $ leads in Eq.(\ref{e-tot}) immediately to $ E_{\mathrm{tot}} (\varphi) = E_{\mathrm{tot}} (\varphi') $ such that the total junction energy is independent of the phase shift $ \varphi $ and the two chiral states are phase decoupled. 

On the other hand, if we assume a four-fold rotation symmetry, then above relations for the energy are only true for $ \alpha = \frac{\pi}{2} (2n+1) $. From this we find
\begin{equation}
E_{\mathrm {tot}} (\varphi) = E_{\mathrm{tot}} (\varphi')=E_{\mathrm{tot}} (\varphi- (2n+1) \pi )
\end{equation}
which is borne out to lowest order by a dependence like $ \cos (2 \varphi) $ as found previously in the GL formulation and is also consistent with the numerical result for the domain wall energy in Fig.~\ref{fig:DomainEnergy}. The absence of phase coupling for the isotropic case means that the critical current through a domain wall along the $c$-axis would vanish and would even be rather small for the anisotropic system.

\section{Half-quantum vortex}
\label{sec:hqv}

We now turn to the question of vortices on a $c$-axis domain wall. As we will show, the $ \pi $-periodicity of the phase shift $ \varphi $ suggests the existence of half quantum vortices (HQV).  The structure and magnetic properties of a single HQV on the domain wall will be discussed in Sec.~\ref{sec:struc}. In Sec.~\ref{sec:sg}, the phenomenology of treating the $c$-axis domain wall as an effective Josephson junction is presented.

\subsection{Structure of the HQV}\label{sec:struc}

For bulk vortices in conventional or one-component superconductors, the phase of the order parameter winds by an integer multiple of $2\pi$ around the singularity at the line defect. For multi-component superconductors, on the other hand, each order parameter component can wind separately, such that more intricate structures of vortices with a fractional magnetic flux are possible\cite{sigrist:1991}.

The $c$-axis domain wall considered in this work supports such fractional vortices, but limited to carrying (integer multiples of) half of a standard flux quantum $ \Phi_0 = hc/2e $. This results from the non-trivial $\pi$-periodicity of the phase shift across the domain wall described in Eq.~(\ref{eq:varphi}) and from the underlying $\cos(2\varphi)$ coupling term in Eq.~(\ref{eq:f_phi}), such that a $\pi$-kink is the smallest possible increase of the phase shift between two stable configurations of the domain wall, unlike the standard Josephson vortices corresponding to a $ 2 \pi $-kink of $ \varphi $. 

\subsubsection{Stability, phase, and shape of the order parameter}

\begin{figure}
\includegraphics[width=0.9\columnwidth]{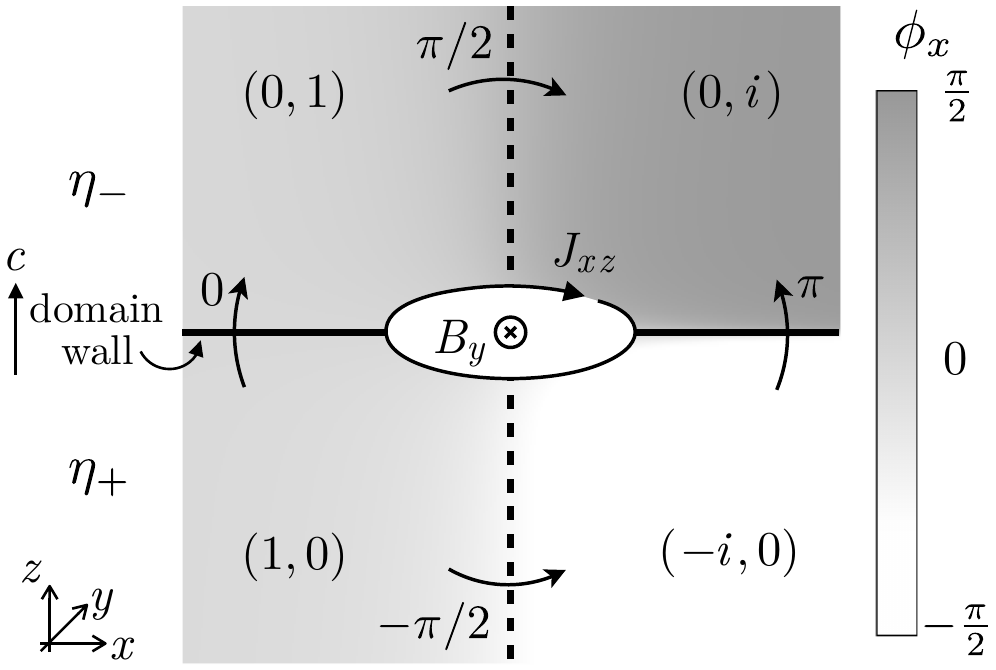}
\caption{Stable setup of a line defect on the $c$-axis domain wall (thick black) with all resulting states of the order parameter $(\eta_+,\eta_-)$ indicated. 
The anisotropy is $\nu<0$, with the allowed phase shifts across the junction of $0$ and $\pi$. Introducing phase shifts $\pm\pi/2$ perpendicular to the domain wall (dashed) connects the two stable configurations far away from the line defect. The magnetic flux line $B_y$ and the circular current $J_{xz}$ are sketched. The background density plot displays the global phase $\phi_x$ at $\nu=-0.11$, also shown in Fig.~\ref{fig:op_hqv}.}
\label{fig:hqv}
\end{figure}

We consider the case of $ \nu < 0 $ where the stable domain walls possess the phase shifts $ \varphi = n \pi $. The $c$-axis domain wall shall be centered at $z=0 $ separating the phases $ \eta_+ $ for $ z < 0 $ and $ \eta_- $ for $z > 0 $. We introduce now a line defect on the domain wall at $ x =0 $ by choosing the phase shift $ \varphi = 0 $ for $ x \to - \infty $ and  $ \varphi = \pi $ for $ x \to + \infty $. As illustrated in Fig.~\ref{fig:hqv}, $ \varphi $ has to change by $\pm\pi/2$ for $z\gtrless 0$ along the $x$-axis to connect these two stable domain wall configurations. Away from the line defect, such a phase gradient does not cost any energy but can be absorbed by the proper gauge $\gamma A_x(x)=\partial_x \varphi(x)$. The resulting states of the order parameter $(\eta_+,\eta_-)$ serving as the boundary conditions are all indicated in Fig.~\ref{fig:hqv}. In addition, the order parameter phase $ \phi_x $ is shown as a density plot in the background, for which we switch back to the basis,
$ {\boldsymbol \eta} = (\eta_x , \eta_y) =  (|\eta_x| e^{i \phi_x}  , |\eta_y| e^{i \phi_y}) $. We also note that the system is still translationally invariant along the $y$-direction.

For the subsequent analysis we neglect any surface effects, that is, we consider an infinite sample. Moreover, for simplicity we assume that to lowest order $A_y=0$, even though this prevents a fully self-consistent analysis because the spatial variation of the order parameter along the $x$-direction is associated with a small but finite $A_y$ component through the self-screening of the induced source current $J_y$ along the $y$-direction. The detailed structure of the magnetic flux pattern around the line defect is computed by minimizing the full GL free energy functional for the boundary conditions indicated in Fig.~\ref{fig:hqv} numerically using a relaxed on-step Newton-Jacobi method, described in detail in Ref.~[\onlinecite{etter:2017}]. To facilitate the computations, a less extreme value of $\gamma_s=10$ is used in this part, while the value of $\gamma_s=20$ as found in the literature for Sr$_2$RuO$_4$ \cite{maeno:2012} was used in the first part.

\begin{figure}
\includegraphics[width=0.9\columnwidth]{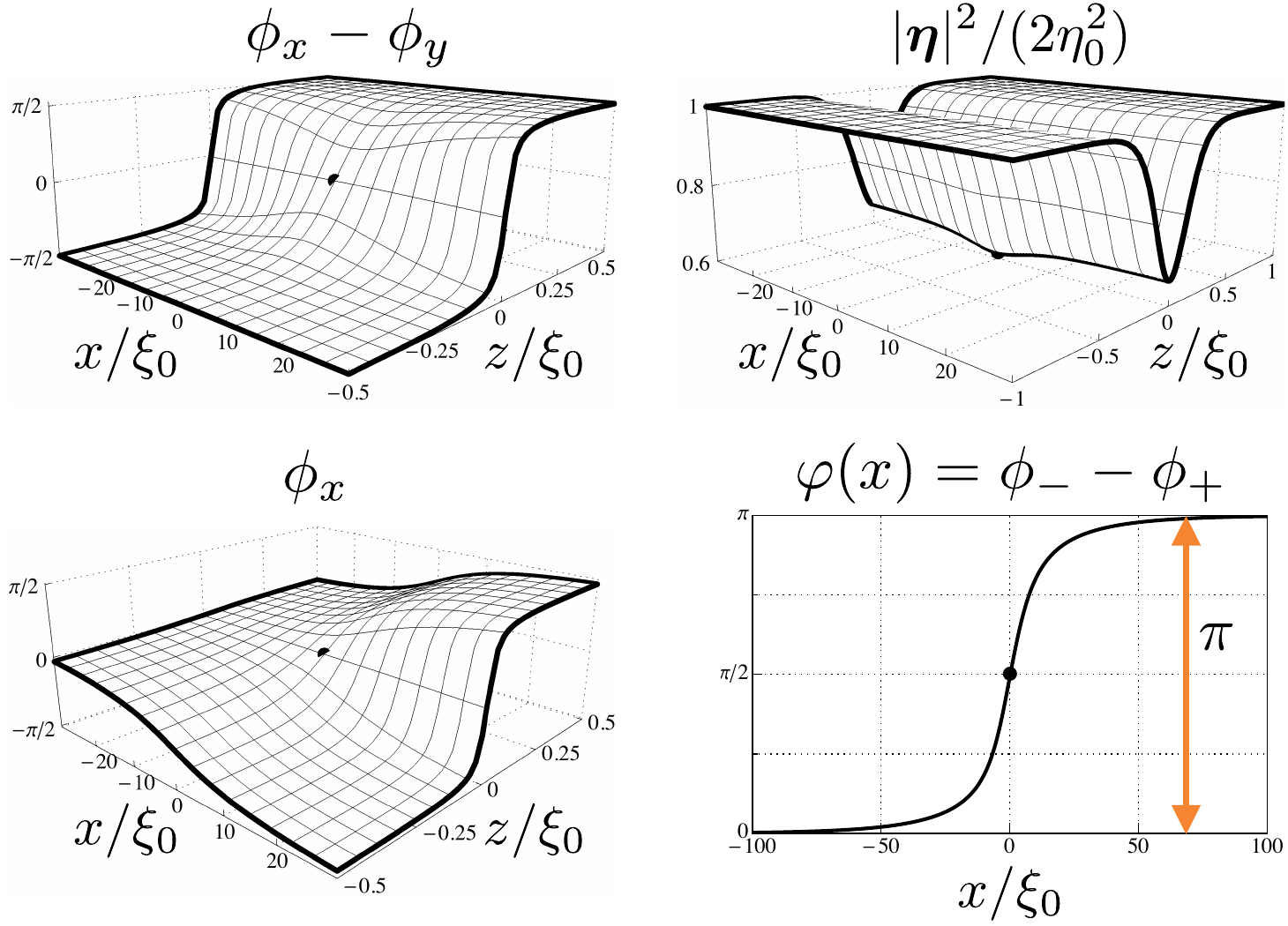}
\caption{Structure of the order parameter around the line defect (black dot) at $(0,0)$ for $\nu=-0.11$. (top left) Phase difference between $\eta_x$ and $\eta_y$, confirming the domain wall. (top right) Total absolute value, suppressed at the domain wall and at the line defect. (bottom left) Global phase, in accordance with the setup; this quantity is shown as the background density plot in Fig.~\ref{fig:hqv}. (bottom right) $\pi$-kink in the phase shift $\varphi(x)$ along the domain wall.}
\label{fig:op_hqv}
\end{figure}

Various quantities extracted from the computational result for the order parameter are shown in Fig.~\ref{fig:op_hqv}, with the position of the line defect at $(x,z)=(0,0)$. The relative phase between the $x$- and the $y$-component of the order parameter is shown in the top left panel. It is $+\pi/2$ for $z>0$ and $-\pi/2$ for $z<0$, in accordance with the setup and consistent with the presence of the domain wall. The absolute value $|\boldsymbol{\eta}|^2$ of the order parameter is shown in the top right panel. As discussed above, it is suppressed at the domain wall. Now, it is additionally reduced at the line defect. The phase $\phi_x=\arg(\eta_x)$ is shown in the bottom left panel and behaves as anticipated in Fig.~\ref{fig:hqv}.
Finally, the panel at the bottom right shows the $\pi$-kink in the phase shift $ \varphi(x) $ along the domain wall, defined here through $\varphi=\phi_--\phi_+$. This quantity is the analogue to the Josephson phase when treating the domain wall as an effective Josephson junction.

\begin{figure}[t]
\centering
\includegraphics[width=0.9\columnwidth]{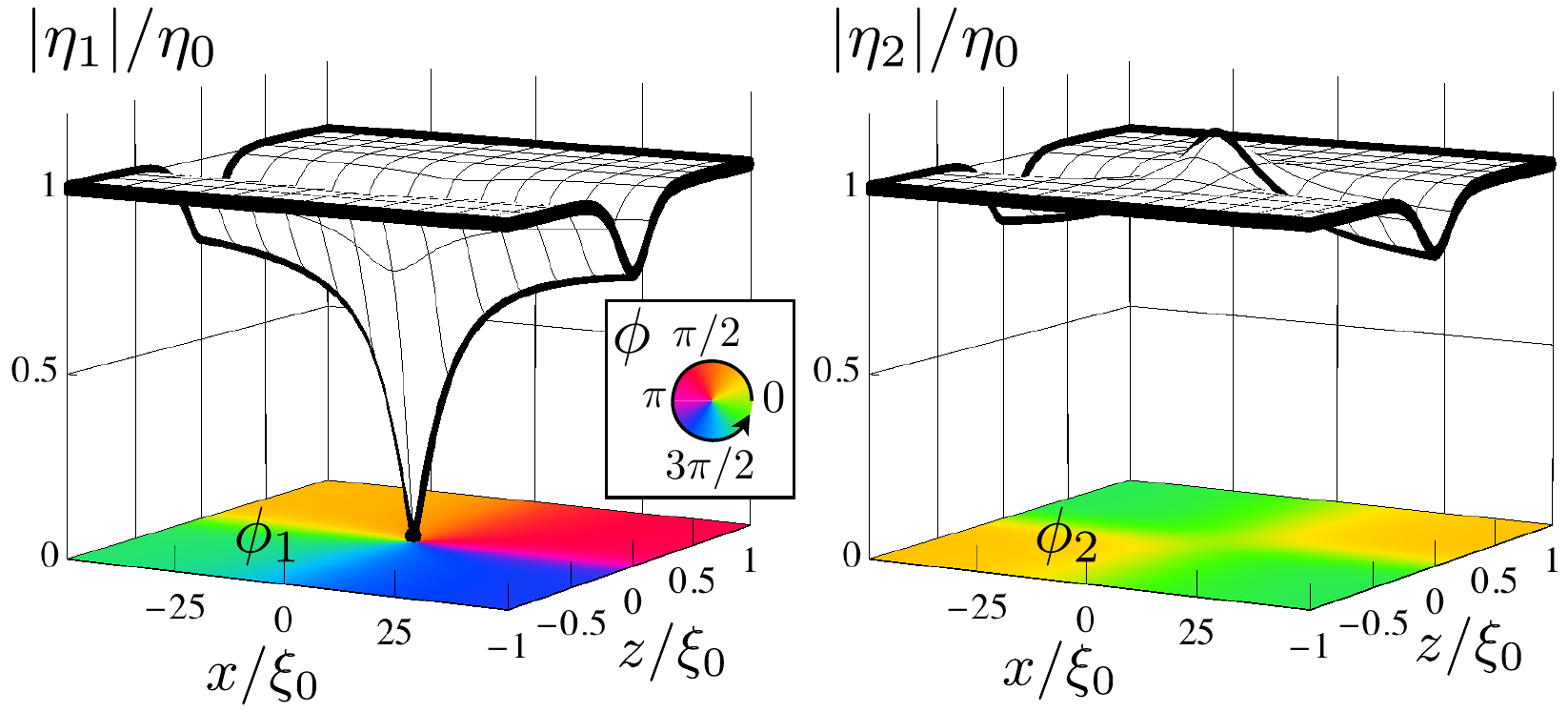}
\caption{The order parameter components $\eta_1$ and $\eta_2$ as given in Eq.~(\ref{eq:singlewind}). The 3D plots show the amplitude and the colored density plots indicate the phase of each component around the line defect at $(0,0)$, all extracted from numerical data for $\nu=-0.11$. The $\eta_1$ component winds by $2\pi$ and is suppressed to zero at the singularity of the line defect. The $\eta_2$ component has no phase winding but is enhanced at the line defect. Both components are suppressed at the domain wall.}
\label{fig:single}
\end{figure}

The line defect is characterized by the winding of one of two order parameter components. In our case this can be extracted by representing the order parameter as
\begin{align}\label{eq:singlewind}
\begin{split}
\boldsymbol{\eta}&=(\eta_1,\eta_2)=\frac{1}{\sqrt{2}}\big(\eta_x-\eta_y,\eta_x+\eta_y\big)\\
&=\frac{1}{\sqrt{2}}\Big((1-i) (\eta_+,\eta_-)+(1+i)(\eta_-,\eta_+)\Big).
\end{split}
\end{align}
The behavior of these two components is illustrated in Fig.~\ref{fig:single} with a three-dimensional plot of their amplitude and a colored density plot of their phase beneath.
While $ \eta_1 $ vanishes at the line defect, $ \eta_2 $ remains finite everywhere. Taking the overall phase structure into account, an unusual vortex carrying half a flux quantum (HQV) emerges from this line defect, as we explicitly show below.

\subsubsection{Characteristic length scales and magnetic properties}

Perpendicular to the domain wall, the magnetic field of the HQV is screened efficiently by inplane currents on the length scale $\lambda_{ab}$. The extension of the magnetic flux distribution along the domain wall, however, depends on the coupling between the two domains and, thus, on the supercurrent which can flow across the domain wall. In order to understand the behavior of the domain wall it is helpful to view it as an effective Josephson junction. The critical current scales like $J_c\propto|\nu|$ for small $|\nu|$, as we will point out in Eq.~(\ref{eq:jc}). The extension of the HQV, like the Josephson vortex, corresponds to the Josephson penetration depth which scales as $\lambda_J\propto1/\sqrt{J_c} \propto1/\sqrt{|\nu|}$. Since the critical current vanishes in the isotropic limit, we also expect that the HQV will dissolve for $ \nu \to 0 $. 
On the other hand, for growing $ | \nu | $ the flux distribution along the $x$-axis shrinks and the picture of the Josephson vortex is not entirely appropriate anymore as new effects come into play.  

For layered superconductors there exist two different screening lengths $\lambda_{ab}$ and $\lambda_c=\gamma_s\lambda_{ab}\gg\lambda_{ab}$ due to screening currents parallel and perpendicular to the layers, respectively, in analogy to the anisotropy of the coherence length  $\gamma_s=\xi_{ab}/\xi_c$. While for conventional Josephson junctions usually $\lambda_J\gg\lambda_\mathrm{London}$, for the situation considered here, the two relevant length scales can be comparable in size already at very small values of $\nu \approx 4 \%$, see Eq.~(\ref{eq:ljapprox}). Once $\lambda_J<\lambda_c$, non-local magnetic properties of the Josephson junction have to be considered, as reviewed in Ref.~[\onlinecite{abdumalikov:2009}]. In this case the long-range screening behavior is more like that of an Abrikosov vortex on the length scale $\lambda_c$, while the core remains Josephson-like, but has a new characteristic length $l=\lambda_J^2/\lambda_c<\lambda_J$\cite{gurevich:1992}. This is derived in detail in Sec.~\ref{sec:sg}, while below we describe how to extract these characteristic length scales from the computational results for the HQV.

\begin{figure}
\includegraphics[width=0.9\columnwidth]{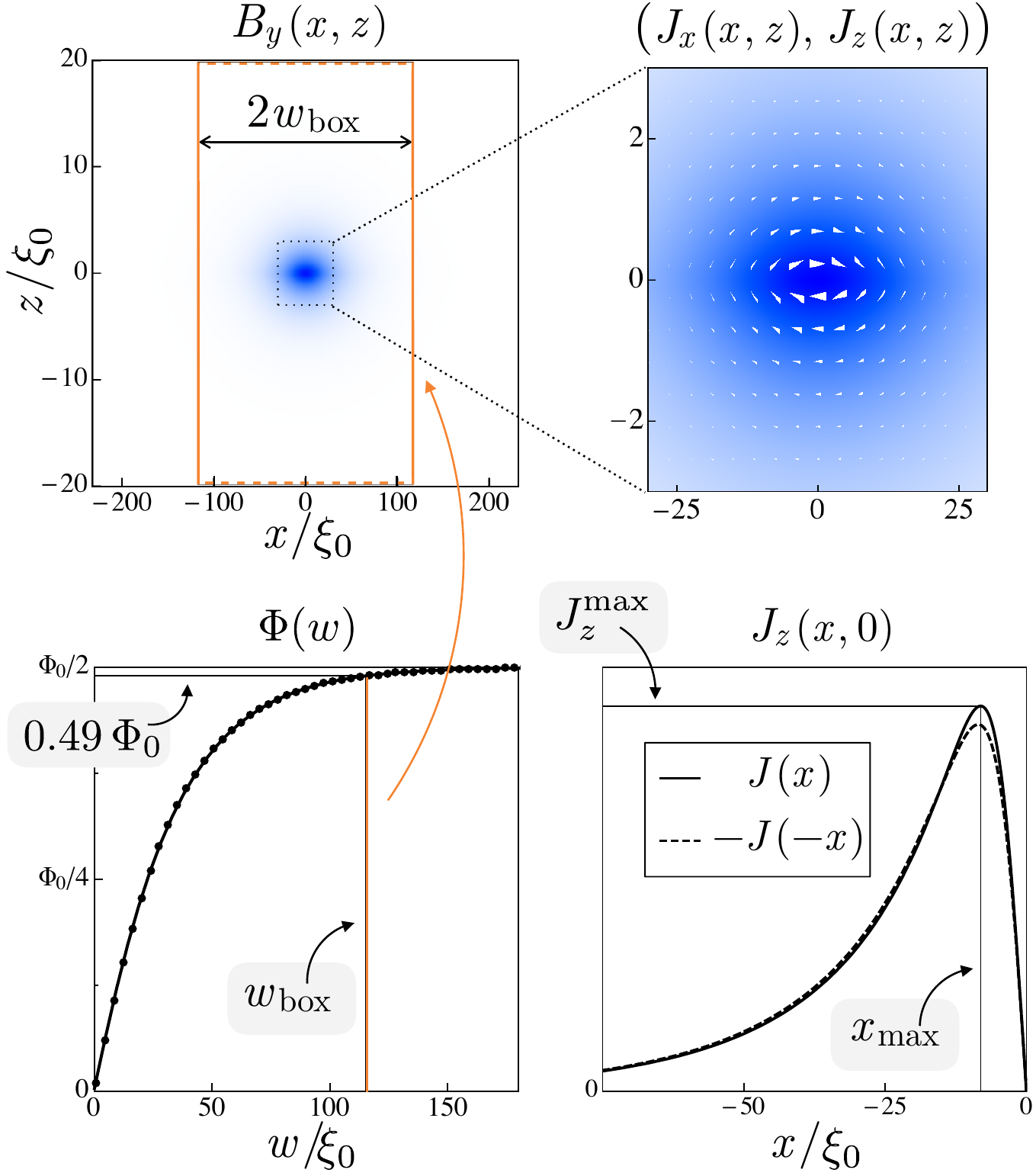}
\caption{Magnetic properties of the HQV for $\nu=-0.11$. (top) Magnetic field $B_y$ with a zoom into the center, also showing the circular current. (bottom left) Total flux $\Phi$ contained in a strip of width $2w$. The width $w_\mathrm{box}$ defined by the box containing $0.49\Phi_0$ is highlighted. (bottom right) Profile of the current $J_z$ across the domain wall (solid) and its inverse (dashed), with the maximum and its position highlighted.}
\label{fig:field_hqv}
\end{figure}

The structure of the magnetic flux line and the current pattern circulating around the HQV are shown in Fig.~\ref{fig:field_hqv}. A density plot of the magnetic field $B_y(x,z)$ is displayed at the top, with an inset zooming in on its center, where a vector plot of the current $(J_x,J_z)$ can be seen. A measure of the extension of the HQV along the $x$-axis can be derived by limiting the integral for the magnetic flux through the boundaries at $x=\pm w$,
\begin{align}
\Phi(w)=\int_{-\infty}^{\infty}\mathrm{d}z\int_{-w}^{w}\mathrm{d}xB_y(x,z),
\end{align}
whose result is shown in the bottom left panel. For $ w \to \infty $ we observe that the flux $ \Phi (w) $ saturates at $ \Phi_0/2 $ as expected for a HQV. 
We now use this behavior to define the length $ w_\mathrm{box} $ through $ \Phi(w_\mathrm{box})=0.49 \Phi_0$ (this somewhat arbitrary cutoff does not qualitatively influence the final results), as indicated in Fig.~\ref{fig:field_hqv}. Furthermore, the core size of the HQV can be estimated using the profile of the current across the domain wall, $J_z(x,0)$, shown in the bottom right panel. The length scale of the core is defined as the position $x_\mathrm{max}$ of the maximal current $J_z^\mathrm{max}$, which in turn gives a measure for the critical current of the domain wall considering it as an effective junction between the two domains. In addition,  $-J_z(-x)$ is shown (dashed), which indicates that the current is slightly asymmetric, while the total current $\int_{-\infty}^{\infty}\mathrm{d}x J_z(x,0)$ still integrates to zero within the numerical accuracy. From a detailed analysis of the full GL free energy functional, and also from symmetry considerations, it becomes apparent that for the spatial variation of the phase shift $\varphi(x)\neq -\varphi(-x)+\pi$. Specifically, the relaxation away from the HQV towards the stable $0$ (or $2\pi$) phase shift occurs on a slightly different length-scale than towards the $\pi$ phase shift, because the inplane gradient terms are fundamentally $2\pi$-periodic only. This small effect, however, does not affect our overall discussion.

\subsection{Junction phenomenology}\label{sec:sg}

We now explore the behavior of the HQV for varying anisotropy $\nu$ by treating the domain wall as an effective Josephson junction. First, the current-phase relation and the critical current are discussed using our variational approach and comparing it to the computational result. Next, a sine-Gordon model is formulated both for the isotropic (small $ \nu$) limit and for the non-local (large $ \nu $) limit. Eventually, the characteristic length scales $w_\mathrm{box}$ and $x_\mathrm{max}$ for the core size and the full size of the HQV as introduced above are analyzed, based on the analytical estimates from the two limits, and compared to the values extracted from the computational results.

\subsubsection{Current-phase relation}

Treating the domain wall as an effective Josephson junction, the Josephson phase difference is associated with the phase shift $\varphi(x)=\phi_--\phi_+$. Based on the $\pi$-periodicity of the phase shift, the current-phase relation behaves to lowest order as
\begin{align}\label{eq:cp0}
J_z(\varphi)=J_c \sin(2\varphi),
\end{align}
with the critical current a parameter to be determined, either from an analytical approach (see below), or extracted from our computational results (see Fig.~\ref{fig:field_hqv}).

Deriving the current-phase relation from the full domain wall free energy self-consistently is beyond the scope of this paper. Instead we focus on the limit of small anisotropies and use the approximation presented in the first part, Eq.~(\ref{eq:dw_en}), assuming a constant total order parameter amplitude $ | \boldsymbol \eta | $ throughout the system and neglecting the inplane gradient coupling terms.
The phase-dependent domain wall free energy is then given by
\begin{align}\label{eq:fedw_p}
\mathcal{F}_\mathrm{dw}(\varphi)&=\frac{3}{\sqrt{2}}b\eta_{b}^4\xi_c\sqrt{1+\nu\cos(2\varphi)}.
\end{align}
In analogy to the Josephson junction, the current density across the domain wall is therefore
\begin{align}\label{eq:cp}
\begin{split}
J_z(\varphi)&=\frac{2\pi c}{\Phi_0}\partial_\varphi \mathcal F_\mathrm{dw} \\
&= \frac{3}{\sqrt{2}}\frac{c\Phi_0}{8(2\pi)^2\xi_{ab}\lambda_{ab}\lambda_c}\frac{-\nu\sin(2\varphi)}{\sqrt{1+\nu\cos(2\varphi)}}.
\end{split}
\end{align}
Approximating for small anisotropies $|\nu|\ll1$, the lowest order expression proposed in Eq.~(\ref{eq:cp0}) is confirmed, with the critical current given by
\begin{align}\label{eq:jc}
J_c\approx  \frac{3}{\sqrt{2}}\frac{c\Phi_0}{8(2\pi)^2\xi_{ab}\lambda_{ab}\lambda_c}|\nu|,
\end{align}
recovering the behavior for the isotropic limit discussed above. When the domains are decoupled, no supercurrent can flow across the domain wall, and indeed $J_c(\nu=0)=0$.

The assumption of a constant total amplitude $|\boldsymbol\eta|^2=|\eta_\mathrm{b}|^2$ was found to be the least valid for small $\nu$, while this is exactly the limit of interest here. We therefore propose a semi-analytical model, where instead of using Eq.~(\ref{eq:fedw_p}) for the domain wall energy directly, the computational result near $\nu\approx0$ is fitted linearly, indicated in Fig.~\ref{fig:fe} by the dashed line. The proportionality $J_c\propto|\nu|$ still holds, but with a different slope, $J_c^\mathrm{fit}\approx 0.5J_c$.

To put the critical current of the domain wall in relation to a situation without any domain wall, i.e. the upper limit, the standard procedure for the depairing current is followed\cite{degennes:1999}. The maximal current along the $c$-direction is then given by
\begin{align}\label{eq:jcdep}
J_{z}^\mathrm{dep}=\frac{8c\gamma K_5|\eta_\mathrm{b}|^2}{3\sqrt{3}\xi_c}=\frac{c\Phi_0}{3\sqrt{3} (2\pi)^2\xi_{ab}\lambda_{ab}\lambda_c}.
\end{align}

\begin{figure}[t]
\centering
\includegraphics[width=0.9\columnwidth]{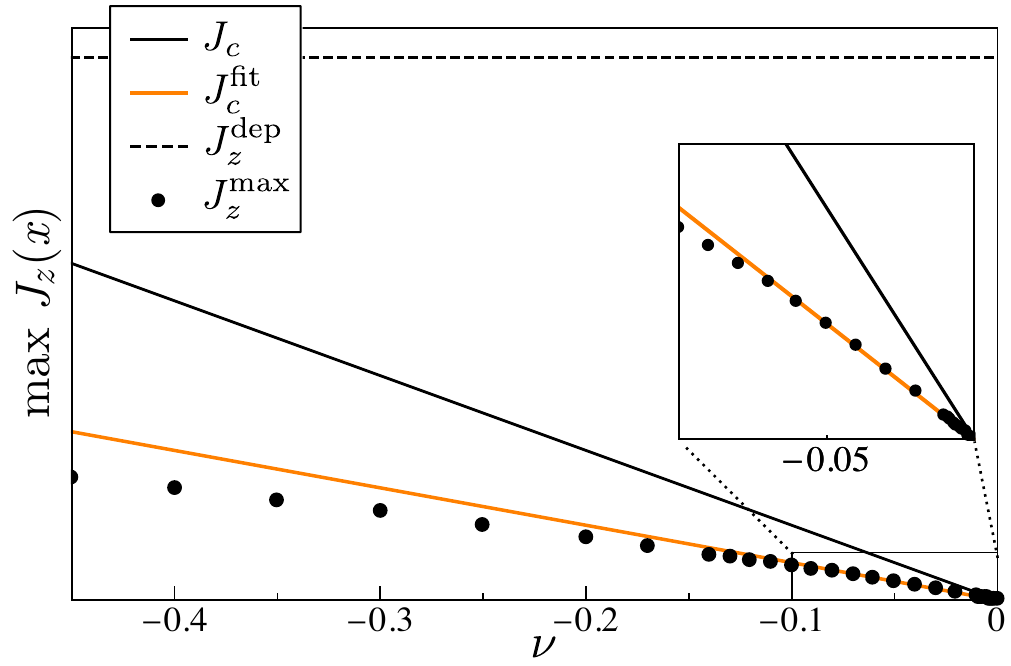}
\caption{Maximum current across the domain wall at the HQV extracted from the computational results (black dots), compared to the critical current obtained from the approximative analytical solution (black line) and from a linear fit to the numerical solution for the free energy (orange). The ultimate limit is the depairing current (dashed). The inset zooms into the limit of small anisotropies.}
\label{fig:jc}
\end{figure}

In Fig.~\ref{fig:jc} we compare the maximal current extracted from the computational results of the HQV $J_z^\mathrm{max}$ (black dots) as defined in Fig.~\ref{fig:field_hqv}; the linear expansion of the critical current $J_c$ from the analytic approximation of the domain wall energy as defined in Eq.~(\ref{eq:jc}) (black line); the critical current $J_c^\mathrm{fit}$ obtained from fitting the numerical result of the domain wall energy as discussed above (orange line); and the depairing current $J_{z}^\mathrm{dep}$ defined in Eq.~(\ref{eq:jcdep}) (dashed line). For small $|\nu|\ll1$ the maximum current of the HQV $J_z^\mathrm{max}$ follows the critical current $J_c^\mathrm{fit}$.
This supports our treatment of the domain wall as an effective Josephson junction and using the lowest order current-phase relation Eq.~(\ref{eq:cp0}) in the limit of small anisotropies.
At higher values of $|\nu|$, the maximum current of the HQV, $J_z^\mathrm{max}$, becomes smaller than the extrapolated value $J_c^\mathrm{fit}$. Here, the lowest order approximation for the current-phase relation at the domain wall is therefore insufficient, and additional effects come into play. 
Indeed, in the limit $|\nu|\rightarrow1$ the domain wall expands along the $c$-axis and spans many layers, such that a simple junction description is no longer warranted.

In the following, the maximal value $J_z^\mathrm{max}$ extracted from the computational results for the HQV will be used as the critical current $J_c$, and the lowest order current-phase relation Eq.~(\ref{eq:cp0}) will be assumed. As we will see below, the crossover to the non-local behavior already happens at a value where this approximation is still valid, such that the structural change of the HQV, because of the different characteristic length scales, can be discussed safely, while an analytical description of the behavior of the HQV at very large anisotropies would require further investigation.

\subsubsection{Nearly isotropic limit}

In the isotropic limit $|\nu|\approx0$, the field of the HQV is locally related to the phase difference at each point within the junction through the standard field-phase relation\cite{orlando:1991}, which for our geometry is given by
\begin{align}\label{eq:sgfieldphase}
B_y(x)=\frac{\Phi_0}{2\pi d}\partial_x\varphi(x),
\end{align}
where $d$ is the effective width of the junction. For the $c$-axis domain wall $d=2\lambda_{ab}$. 
Together with the lowest order current-phase relation Eq.~(\ref{eq:cp0}), this results in the sine-Gordon equation
\begin{align}
J_c\sin(2\varphi)=\frac{c}{4\pi}\frac{\Phi_0}{2\pi d} \partial^2_{x}\varphi(x),
\end{align}
which has the solution
\begin{align}\label{eq:sgphase}
\varphi(x)=2\arctan\left(e^{x/\lambda_J}\right),
\end{align}
with Josephson penetration depth $\lambda_J$ given by
\begin{align}\label{eq:lj}
\lambda_J=\sqrt{\frac{c\Phi_0}{16\pi^2 d J_c}}.
\end{align}
The resulting current has a maximum value $J_z^\mathrm{max}=J_c$ by construction and at the position $x_\mathrm{max}=\mathrm{arcsinh}(1)\lambda_J\equiv\tilde\lambda_J$. This is the single characteristic length measuring the extension of the HQV in the isotropic (local) limit.

Using the approximative expression for the critical current discussed above, $J_c^\mathrm{fit}$, the Josephson penetration depth in the isotropic limit is given by
\begin{align}\label{eq:ljapprox}
\lambda_J\approx\sqrt{\frac{2\sqrt{2}}{3\kappa\gamma_s|\nu|}}\lambda_c.
\end{align}
The Josephson penetration depth and the relevant screening length coincide, i.e. $\lambda_J\approx\lambda_c$, at $\nu_\mathrm{cross} \approx 4\%$, where the lowest order current-phase relation is certainly still valid, see Fig.~\ref{fig:fe} and Fig.~\ref{fig:jc}. For higher values of $|\nu|$, where $\lambda_J<\lambda_c$, the standard sine-Gordon model is no longer valid, as $\lambda_c$ provides the fundamental screening length. We conclude that the non-local effects already come into play for very small anisotropies and can be discussed using the lowest order current-phase relation.

\subsubsection{Non-local limit}

When $\lambda_J<\lambda_c$, the field of the HQV depends on the phase difference at all points of the junction through a case-specific non-locality kernel $G(x,z,x')$ as
\begin{align}
B_y(x,z)=\frac{\Phi_0}{2\pi}\int_{-\infty}^{\infty}G(x,z,x')\partial_{x'}\varphi(x')\mathrm{d}x',
\end{align}
as discussed in the review Ref.~[\onlinecite{abdumalikov:2009}].
The local case is recovered by $G(x,z,x')=\delta(x-x')/d$. Since the lowest order current-phase relation holds beyond $\nu_\mathrm{cross}$, we propose the following non-local sine-Gordon model,
\begin{align}\label{eq:nlsg}
J_c\sin(2\varphi)=\frac{c}{4\pi}\frac{\Phi_0}{2\pi}\partial_x\int_{-\infty}^{\infty}G(x,z,x')\partial_{x'}\varphi(x')\mathrm{d}x'.
\end{align}
The kernel is taken as for the case of vortices in layered superconductors with planar defects as described in Ref.~[\onlinecite{gurevich:1996}], given by a modified Bessel function
\begin{align}
G(x,z,x')=\frac{1}{2\pi\lambda_{ab} \lambda_c}K_0\left(\sqrt{\frac{(x-x')^2}{\lambda_c^2}+\frac{z^2}{\lambda_{ab}^2}}\right).
\end{align}
The field of the HQV can then approximately be written as (see Eq.~(34) in Ref.~[\onlinecite{gurevich:1996}])
\begin{align}\label{eq:nonlocalfield}
B_y(x,z)=\frac{\Phi_0}{4\pi\lambda_c\lambda_{ab}} K_0\left(\sqrt{\frac{x^2}{\lambda_c^2}+\left(\frac{l}{\lambda_c}+\frac{|z|}{\lambda_{ab}}\right)^2}\right),
\end{align}
with the characteristic length scale $l=\lambda_J^2/\lambda_c$
\begin{align}\label{eq:lengthl}
l=\frac{c\Phi_0}{32\pi^2\lambda_{ab}\lambda_c J_c}.
\end{align}
Note that this form for the field, Eq.~(\ref{eq:nonlocalfield}), gives the correct asymptotics within the present approach, but is not normalized correctly due to the core cut-off length $l$.
The maximal current is still $J_c$ by construction, but the peak position is now at $x_\mathrm{max}=l$, such that $l$ can be considered the core size of the HQV. The long-range behavior is Abrikosov-like and determined by $\lambda_c$.
Using for the critical current again $J_c^\mathrm{fit}$ as for Eq.~(\ref{eq:ljapprox}), this results in
\begin{align}\label{eq:lengthl2}
l\approx\frac{2\sqrt{2}}{ 3|\nu|}\xi_{ab}=\frac{2\sqrt{2}}{ 3\kappa\gamma_s|\nu|}\lambda_c,
\end{align}
such that even $l\ll\lambda_c$ for large anisotropies $|\nu|\gg\nu_\mathrm{cross}$.

There are now two important length scales of very different size. Therefore, attaining a proper resolution of the core of the HQV while covering a large enough system size to accommodate the full HQV poses a challenge in the numerical minimization. This can be solved by using a fixed but high number of mesh points while adapting the step size between the mesh points with changing $\nu$ (for details see Ref.~[\onlinecite{etter:2017}]).

\subsubsection{Results and discussion}

Finally, the expressions for the characteristic length scales, as derived above, in both the isotropic and the non-local limit will be compared with the computational results to examine the overall structural behavior of the HQV on the $c$-axis domain wall.

\begin{figure}
\includegraphics[width=0.9\columnwidth]{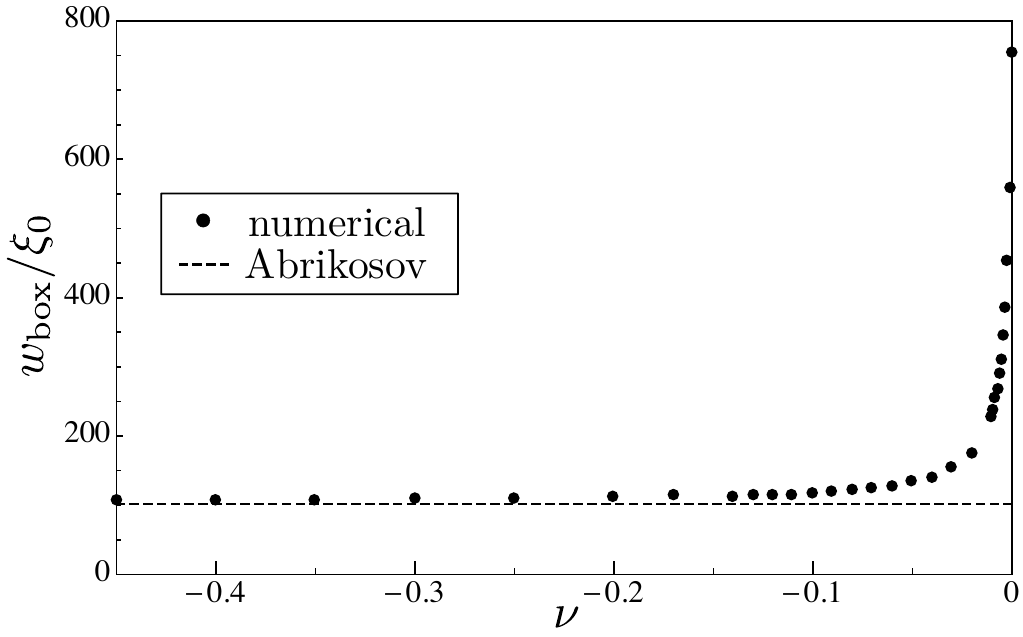}
\caption{Width $w$ of the box containing $0.49\Phi_0$ as a function of the anisotropy $\nu$ extracted from computational results (black dots) with the lower limit given by the strip width for an anisotropic Abrikosov vortex (dashed line) with screening lengths $\lambda_c$ and $\lambda_{ab}$.}
\label{fig:wbox}
\end{figure}

First we address the effective magnetic screening length along the domain wall which we estimated through our definition of $w_\mathrm{box}$ (Fig.~\ref{fig:field_hqv}). 
In the range of larger $ |\nu| $ the long-distance behavior can be well approximated by an Abrikosov type of vortex of an anisotropic superconductor \cite{kogan:1981,clem:1990},
whose field is given by
\begin{align}
B_y(x,z)=\frac{\Phi_0}{4\pi\lambda_c\lambda_{ab}} K_0\left(\sqrt{\frac{x^2}{\lambda_c^2}+\frac{z^2}{\lambda_{ab}^2}}\right),
\end{align}
ignoring the core region. Fig.~\ref{fig:wbox} shows the value of $w_\mathrm{box}$ extracted from the numerical results (black dots) and the limiting $w_\mathrm{A}$ computed from the above expression for the Abrikosov vortex (dashed line), where only the screening lengths $\lambda_{ab}$ and $\lambda_c$ enter. 
We find that for $|\nu|\approx0$ the HQV expands as predicted, indicating that the critical current vanishes in the isotropic limit. On the other hand, for growing $ | \nu | $ the vortex size is  well described by the long-distance behavior of an Abrikosov-type vortex. 

\begin{figure}
\includegraphics[width=0.9\columnwidth]{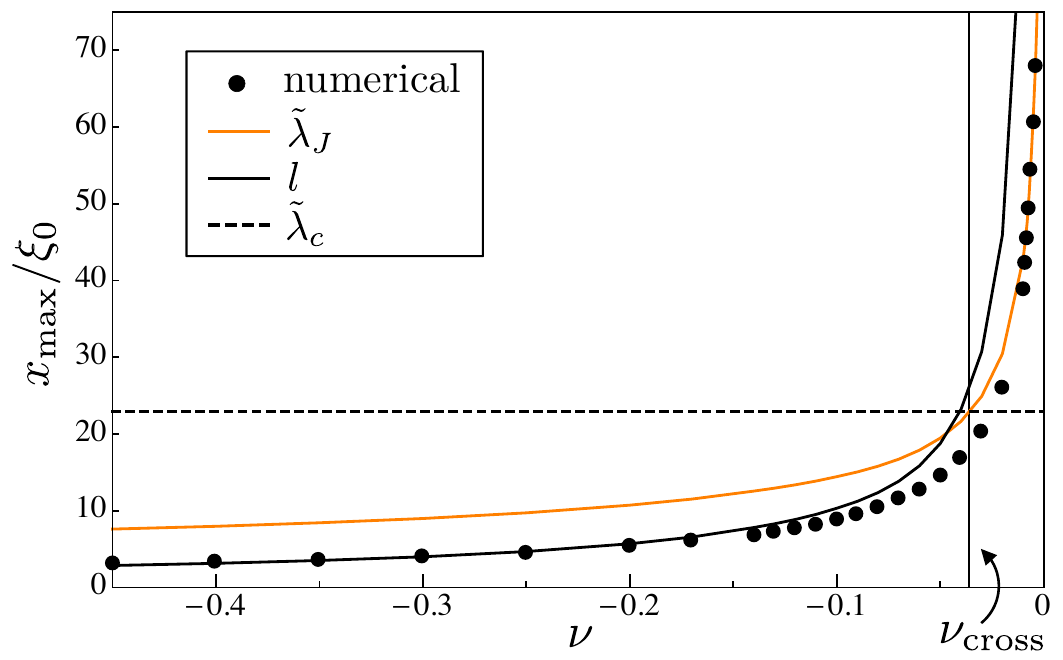}
\caption{
Position of the maximum current across the domain wall at the HQV, extracted from the numerical results (black dots), compared to $\tilde\lambda_J$ (orange) for the isotropic limit $|\nu|\approx0$ and to $l$ (black) for the non-local limit.
The fundamental screening length $\tilde\lambda_c$ (dashed) and the crossover anisotropy $\nu_\mathrm{cross}$ are also indicated.}
\label{fig:xmax}
\end{figure}

Let us turn to the core size $x_\mathrm{max}$ of the HQV, which we defined as the position of the maximum current across the domain wall (see Fig.~\ref{fig:field_hqv}). In Fig.~\ref{fig:xmax} the computational results for $x_\mathrm{max}$ (black dots) are shown, together with the position $\tilde\lambda_J$ defined via the Josephson penetration depth ($\tilde\lambda_J=\mathrm{arcsinh}(1)\lambda_J)$ (orange line), and the characteristic length of the non-local limit $l$ (black line). For both, the critical current used is extracted from $J_z^\mathrm{max}$, and inserted into Eq.~(\ref{eq:lj}) and Eq.~(\ref{eq:lengthl}), respectively. In addition, the relevant screening length $\tilde\lambda_c=\mathrm{arcsinh}(1)\lambda_c$ is displayed (dashed line), as well as the value $\nu_\mathrm{cross}$ of the anisotropy at which $\tilde\lambda_J=\tilde\lambda_c$. 
The size of the vortex core in the isotropic limit $|\nu|\approx0$ is indeed given by $\lambda_J$, while it is determined by $l$ for larger anisotropies, and changes in the crossover region around $\nu_\mathrm{cross}$.

The variation of the extension of both the magnetic flux distribution and the core size are captured well within our sine-Gordon models. Restricting to lowest order current-phase relation gives consistent results valid even beyond the crossover region. Only at larger anisotropies bulk effects start to interfere and the HQV develops a normal core. While with increasing $|\nu|$ the order parameter amplitude $ | \boldsymbol \eta |$ becomes less reduced at the domain wall (see Fig.~\ref{fig:fe}), it actually shrinks more strongly at the center of the HQV. In this regime, we observe a substantial deviation of the actual critical current relative to the approximation $J_c^\mathrm{fit}$ such that the simple junction description is therefore no longer valid.

\section{Conclusion}

In this paper, the structure of $c$-axis domain walls in chiral $p$-wave superconductors and the formation of HQVs has been investigated. The main results are that the coupling between the chiral domains across the domain walls is weak and vanishes completely if the electronic band structure near the Fermi surface is isotropic, and that $c$-axis domain walls can host flux lines carrying only half of a standard flux quantum. These flux lines are shown to dissolve in the limit of an isotropic system. Both features are connected with a reduced critical current through these domain walls, as the coupling is weak and phase slips limiting the supercurrent flow are cheap. 

The possibility that $c$-axis domain walls introduce a severe reduction of the critical current along the $c$-axis can have interesting experimentally testable implications. 
Imagine a sample of a chiral $p$-wave superconductor which is rather narrow for the inplane directions but long along the $c$-axis. For such a sample the critical current along the 
$c$-direction would be strongly dependent on the cooling history, i.e. whether  domain walls are realized or not. For the given geometry  $c$-axis domain walls are 
more likely realized in a fast cooling process. On the other hand, a likely domain-free phase could be realized for slow cooling in a small $c$-axis oriented field.
Evidence for the described impact of domain walls would be a consistent strongly history dependent magnitude of the critical current measured along the $c$-axis.
Moreover, this would be further support for the realization of chiral superconductivity in a material like Sr$_2$RuO$_4$. A similar phenomenology could be expected  in
URu$_2$Si$_2$, a candidate for chiral $d$-wave pairing of the type $ d_{xz} \pm i d_{yz} $, and in many respects similar to the chiral $p$-wave phase\cite{kasahara:2007,schemm:2015}. 

We can further extend our discussion to higher-angular-momentum chiral states such as the chiral $d$-wave state whose gap function has the generic form $ \psi(\boldsymbol k) = \Delta_0 (k_x \pm i k_y)^2 $ as is proposed for the quasi-two-dimensional system SrPtAs\cite{nishikubo:2011,biswas:2013,fischer:2014} and for heavily doped graphene\cite{nandkishore:2012}. Here, the angular momentum  in the isotropic limit is $ J_z = \pm 2 \hbar$. The analogous analysis yields a strong suppression of the phase coupling and a current-phase relation of the $c$-axis domain as $ \cos (4 \varphi) $, such that the domain wall vortices carry quarter quanta of the standard flux quantum. 

\begin{acknowledgments}
We would like to thank D.~Agterberg, A.~Bouhon, M.~Fischer, D.~Geshkenbein, and C. Kallin for helpful discussions. SBE and MS are supported by a grant of the Swiss National Science Foundation (No. 163186 and 184739). WH is grateful for the hospitality of the Pauli Center for Theoretical Studies at ETH Zurich.
\end{acknowledgments}

\section*{Contributions}
All authors were involved in the concept, design and interpretation of the research and the preparation of the manuscript. SBE executed the analytical and numerical analysis of Secs.~\ref{sec:gl} and \ref{sec:hqv}; WH executed the analytical and numerical analysis of Sec.~\ref{subsec:BdG}.

\bibliography{cDomain_REF}

\end{document}